\documentclass[12pt]{iopart}

\usepackage{graphicx}
\usepackage{bm}
\usepackage{epstopdf}
\usepackage{amsfonts}
\usepackage{latexsym,enumerate}
\usepackage{epsfig}
\usepackage[colorlinks=true,linkcolor=blue,urlcolor=blue,citecolor=blue]{hyperref}
\usepackage[T1]{fontenc}
\usepackage[latin9]{inputenc}
\usepackage{hyperref}
\hypersetup{
    colorlinks,
    citecolor=red,
    filecolor=green,
    linkcolor=blue,
    urlcolor=brown
}
\usepackage{iopams}
\hypersetup{linktocpage}
\usepackage{amssymb}
\usepackage{amsthm}
\usepackage[lofdepth,lotdepth]{subfig}

\newcommand{\ket}[1]{\left\vert#1\right\rangle}
\newcommand{\bra}[1]{\left\langle#1\right\vert}

\newcommand{\half}{\mbox{$\textstyle \frac{1}{2}$} }
\newcommand{\rmc}{\mathrm{c}}

\newcommand{\rmh}{\mathrm{h}}

\newcommand{\rms}{\mathrm{s}}
\newcommand{\rmw}{\mathrm{w}}
\newcommand{\rmx}{\mathrm{x}}
\newcommand{\rmy}{\mathrm{y}}
\newcommand{\rmz}{\mathrm{z}}
\newcommand{\rmcc}{\mathrm{cc}}
\newcommand{\rmee}{\mathrm{ee}}
\newcommand{\rmhh}{\mathrm{hh}}
\newcommand{\rmsc}{\mathrm{sc}}
\newcommand{\rmse}{\mathrm{se}}
\newcommand{\rmin}{\mathrm{in}}

\newcommand{\rmout}{\mathrm{out}}
\newcommand{\rmcsh}{\mathrm{csh}}
\newcommand{\rmth}{\mathrm{th}}

\theoremstyle{definition}

\def\cc{{\cal C}}

\def\ch{{\cal H}}
\def\cn{{\cal N}}

\def\ct{{\cal T}}

\newcommand{\id}{\mathbb{I}}

\def\beq{\begin{equation}}
\def\eeq{\end{equation}}
\def\bearr{\begin{eqnarray}}
\def\eearr{\end{eqnarray}}
\def\beal{\begin{equation}\begin{array}{ll}}
\def\eeal{\end{array}\end{equation}}

\begin{document}
\title{An out-of-equilibrium non-Markovian Quantum Heat Engine}
\author{Marco Pezzutto$^{1,2}$, Mauro Paternostro$^3$, and Yasser Omar$^{1,2}$}
\address{$^1$ Instituto de Telecomunica\c{c}\~{o}es, Physics of Information and Quantum Technologies Group, Lisbon, Portugal}
\address{$^2$ Instituto Superior T\'ecnico, Universidade de Lisboa, Portugal}
\address{$^3$ Centre for Theoretical Atomic, Molecular and Optical
Physics, School of Mathematics and Physics, Queen's University
Belfast BT7 1NN, United Kingdom}
\ead{marco.pezzutto@tecnico.ulisboa.pt}

\begin{abstract}
We study the performance of a quantum Otto cycle using a harmonic work medium and undergoing collisional dynamics with finite-size reservoirs. We span the dynamical regimes of the work strokes from strongly non-adiabatic to quasi-static conditions, and address the effects that non-Markovianity of the open-system dynamics of the work medium can have on the efficiency of the thermal machine. While such efficiency never surpasses the classical upper bound valid for finite-time stochastic engines, the behaviour of the engine shows clear-cut effects induced by both the finiteness of the evolution time, and the memory-bearing character of the system-environment evolution.\end{abstract}


\section{Introduction}
\label{sec:intro}

The study of work- and heat-exchanges at the quantum scale~\cite{Talkner07,Gallego16,Goold15} is paving the way to the understanding of how quantum fluctuations influence the energetics of non-equilibrium quantum processes. In turn, such fundamental progress is expected to have significant repercussions on the design and functioning of quantum heat machines~\cite{Quan07,Ghosh18,Seah18,Esposito09,Gelbwaser13,Gardas15}.

Such devices thus play the role of workhorses for the explorations of the potential advantages stemming from the exploitation of quantum resources for thermodynamic applications at the nano-scale~\cite{Francica2017, Watanabe2017}. Theoretical models of microscopic heat engines based on the use of working medium comprising two-level systems~\cite{Linded10} or quantum harmonic oscillators~\cite{Kosloff17} have been introduced. Such designs appear increasingly close to grasp in light of the recent progresses in the experimental management of (so far classical) thermal engines using individual particles~\cite{Abah12,Rossnagel16} or mechanical systems~\cite{Zhang14a,Zhang14b,Dong15}.

Is it possible to pinpoint genuine signatures of quantum behaviour that influence the thermodynamics of a system in ways that could never be produced by a classical mechanism~\cite{Gelbwaser18}? How would quantum mechanics enhance the performance of a quantum thermal engine beyond anything achievable classically~\cite{Abah14,Niedenzu16,Campisi15,Wright18,Tercas17,Francica17}? Do coherences in the energy eigenbasis~\cite{Scully03,Brunner14,Binder15,Hardal15} or non-thermal reservoirs~\cite{Rossnagel14,Alicki15,XYZhang14}, such as those employing squeezing~\cite{Huang12,Long15}, represent exploitable (quantum) thermodynamic resources? 
The question whether quantum non-Markovianity may constitute an exploitable thermodynamic resource is also object of intense studies: in~\cite{Uzdin16} it is shown that quantum heat machines equivalence, valid in the limit of small actions, can be extended to the non-Markovian regime; in~\cite{Gelbwaser13,Bylicka16} non-Markovianity is shown to enhance work extraction by erasure, exploiting system-environment correlations when the thermodynamic cycle duration is below the reservoir memory time; in~\cite{Thomas18} the thermodynamics of interaction with non-Markovian reservoirs is analized, confirming that work extraction can be enhanced by non-Markovian reservoirs, but also showing that, once a minimum cost for non Markovianity is taken into account, the second law retains its validity, and that an Otto cycle with non-Markovian reservoirs can be mapped to a Carnot cycle with Markovian reservoirs. 

In this paper, we contribute to the ongoing quest for the formulation of a fully quantum framework for thermodynamics by studying the finite-time performance of a heat engine opera\-ting an Otto cycle whose working medium is a quantum harmonic oscillator. Hot and cold environments are model\-led via a collections of spin-1/2 particles (\Fref{fig1}). The work strokes of the cycle are implemented via parametric changes of the frequency of the harmonic oscillator, while heat exchanges result from {\it collisional} dynamics with the environments that may allow for memory effects~\cite{Ciccarello17}. The significant flexibility and richness of dynamical conditions of collisional models is perfectly suited to the exploration of non-Markovian dynamics in a wide range of conditions~\cite{Ciccarello13,McCloskey14,Lorenzo15,Kretschmer16,Pezzutto16,Cakmak17,Lorenzo17,Campbell18}. 
The scope of our study is twofold: on the one hand, we investigate work transformations of controlled yet varia\-ble duration, spanning the whole range from an infinitely slow (and thus adiabatic) transformations, to the opposite extreme of a sudden quench. On the other hand, by including intra-environment interactions, we allow for the emergence of memory effects and thus non-Markovianity in the dynamics of the engine. We investigate numerically the behaviour of the engine and its performance in the two cross-overs from adiabaticity to sudden quench, and from Markovianity to non-Markovianity. We aim at identifying the optimal trade-off between efficiency and speed, and the role and impact of memory effects on the engine performance. 

Among the results reported in this paper is the demonstration that the efficiency of the device always decreases as we approach the sudden-quench regime, and the quantification of an optimal time at which the power output is maximum. We complement these results with a study of the irreversibility as measured by the irreversible work. Intra-environment interactions, in turn, seem to have no effect on the long-time engine performance. However, they affect the transient of the evolution of the engine by lowering the efficiency of the heat-transfer process -- at least in the case when the both the engine and the environment particles are initialized in a thermal state. In no case we observe a performance exceeding the classical bounds, which is in agreement with the result reported in~\cite{Gelbwaser18}. We do observe however a strong connection between the detection of non-Markovianity and the coherences in the initial engine state. Finally, the analysis of the behaviour of the machine at different temperatures allowed us to single out the parameter regime in which it behaves as a refrigerator rather then a thermal engine.

The remainder of the article is structured as follows. Sec.~\ref{sec:model} introduces our model for heat engine, describing how the work and heat transformations are realized. Sec.~\ref{sec:results} presents the results of our quantitative analysis, while in Sec.~\ref{sec:conc} we draw our conclusions.

\section{The engine model}
\label{sec:model}

\begin{figure}[t]
\begin{center}
{{\bf (a)}\hskip7cm{\bf (b)}}\\
\includegraphics[width=0.5\columnwidth]{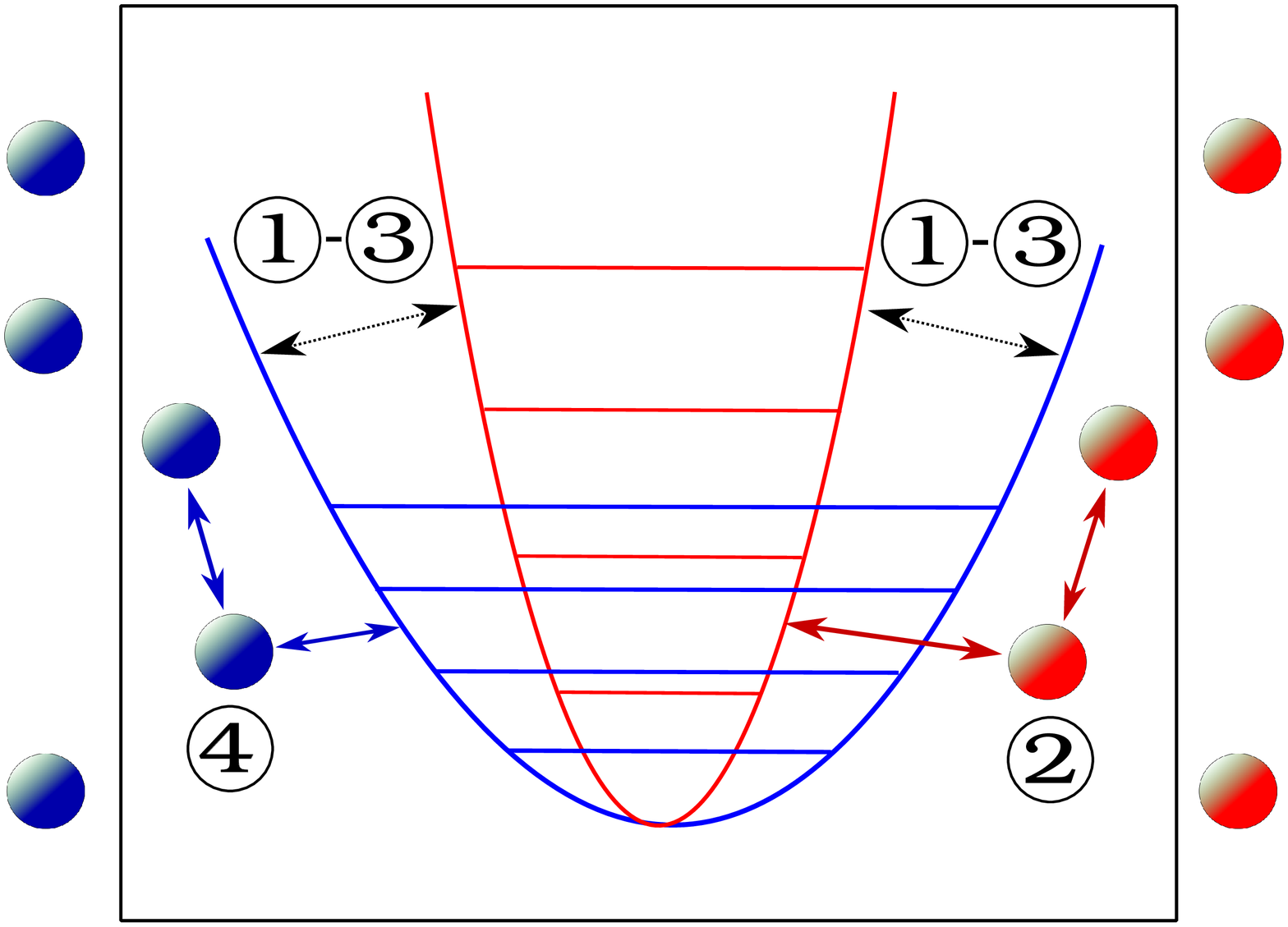}~~~~
{\includegraphics[width=0.4\columnwidth]{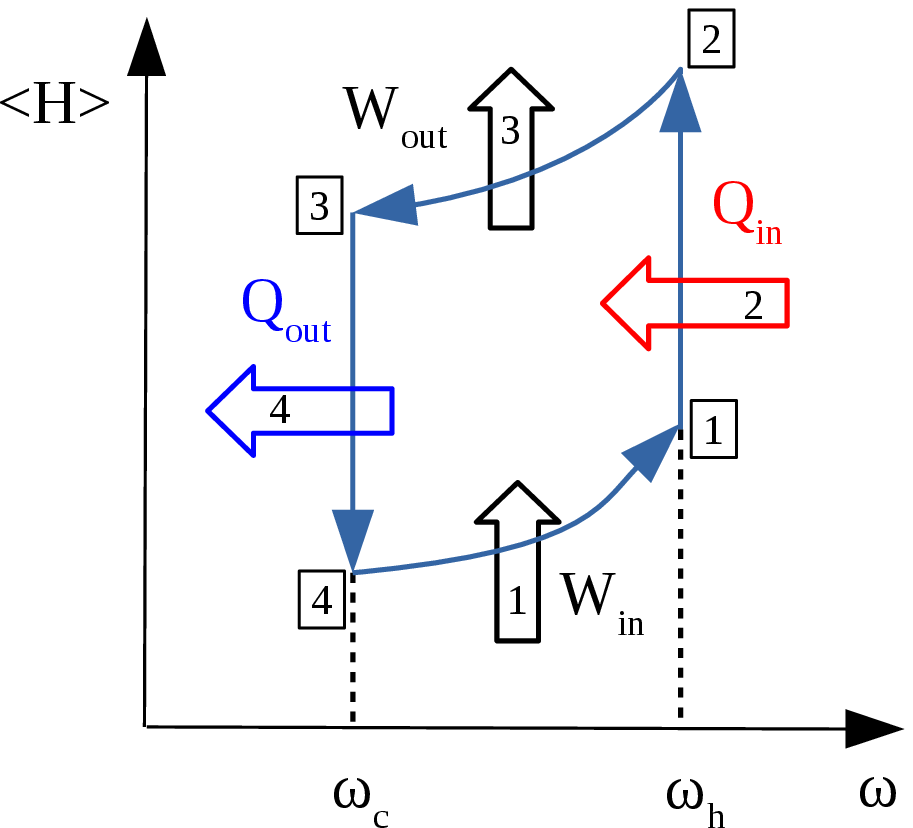}}
\end{center}
\caption{{\bf (a)} We study an engine performing an Otto cycle with a quantum harmonic oscillator as the working medium, which in turn interacts with two environments composed of spin-$1/2$ particles with energy spacing $\omega_c$ and $\omega_h$ respectively. Work is done on/by the oscillator by changing the frequency of its potential between the two extremes $\omega_c$ and $\omega_h$, while in isolation from the environments (cf. Sec.~\ref{sec:work}). Heat is exchanged with the latter through collisions with the spin-$1/2$ particles (cf. Sec.~\ref{sec:heat_coll}). Additional intra-environment interactions allow the environments to keep memory of past interactions with the engine. 
{\bf (b)} As its classical version, the cycle is composed by four strokes: two isentropic (strokes 1 and 3), where work is performed on or by the engine, and two (strokes 2 and 4), during which heat is exchanged with the reservoirs. In our model, the control parameter is the oscillator frequency $\omega$, whose changes play the role of an effective modification in volume in the classical version of the engine. Therefore, strokes 2 and 4 are analogous to isochoric transformations. On the vertical axis, we report the average internal energy of the oscillator $\langle H \rangle$, which quantifies the energy exchanges resulting from the four strokes.}
\label{fig1}
\end{figure}

We study a model of heat engine operating according to an Otto cycle, whose working medium is a quantum harmonic oscillator governed by the Hamiltonian
\beq
\label{eq:HO_Hamiltonian}
{H}_{\rms}(t)= \frac{{p}^2}{2m} + \frac{m \omega^2(t) {x}^2}{2}.
\eeq
The subscript "s" stands for "system" as we may regard the engine as our main system of interest.
The Otto cycle consists of two work strokes and two heat strokes. 
The work strokes are implemented by changing the frequency $\omega$ of the harmonic potential. The hot and cold environments are modelled as a collection of spin-1/2 particles with Hamiltonian
\beq
{H}^{(n)}_{\rme} = \frac{1}{2} \hbar \omega_{\rme } {\sigma}^{\rmz}_{\rme,n }, \qquad \omega_{\rme } >0, \qquad \rme = \rmc,\rmh
\eeq
for the $n^\mathrm{th}$ particle. The subscript h (c) labels a particle in a hot (cold) reservoir.
The working medium interacts with them through a {collisional model}, similar to the one employed in \cite{Pezzutto16}. The details of these dynamical processes, pictured in Figure~\ref{fig1}, are outlined in following Subsections.

\subsection{Details of the cycle operation and thermodynamics of the process}
\label{sec:cycle}

We now outline the protocol through which the Otto cycle is implemented, and the thermodynamic quantities that will be central to our analysis. We start with the {internal energy of the working medium}
	\beq
	\label{eq:IntEn}
	E:=\Tr[\rho_{\rms} {H}_{\rms}].
	\eeq
The second quantity of relevance is the {\it work} done on/by the engine during a work-producing stroke. As no heat is exchanged in one of such strokes, the difference between the values of the internal energy of the engine at the initial and final points of the stroke quantifies the exchanged work. We thus have
	\beq
	\label{eq:work}
	W:=E^{(k)}_{\mathrm{in}}-E^{(k)}_{\mathrm{fin}},
	\eeq
    where $k=1,3$ identifies the work-producing strokes.
In what follows, we use the usual convention that $W>0$ when work is \emph{performed by} the engine. This is also in agreement with a definition of the average exchanged work based on the so-called two-projective-measurement approach~\cite{Deffner08}.

Similarly to the above considerations, no work is exchanged during a heat-exchanging stroke, so that the difference between the values of the internal energy of the engine at the initial and final points of the stroke provides an estimate of the exchanged {\it heat} $Q$. Therefore
	\beq
	\label{eq:heat}
	Q:=E^{(k)}_{\mathrm{fin}}-E^{(k)}_{\mathrm{in}},
	\eeq
where $Q>0$ if it is \emph{absorbed} by the work medium, and $k=2,4$ is the label for the heat-producing strokes. An engine-environment interaction that conserves the total energy [such as the one illustrated in  Sec.~\ref{sec:heat_coll}], is a physi\-cally sound description of a heat transfer process, as it is well suited to describe the heat exchange as a flow of energy from one system (engine or environment) to the other. Moreover, it is consistent with a more general definition of the exchanged heat as the \emph{difference of the environment internal energy}.    

The environmental particles are assumed to be all prepared in a single-particle thermal state, 
\beq
\rho^{(n)}_{\rme} = \frac{e^{-\beta_{\rme} {H}^{(n)}_{\rme}}}{\Tr \,[e^{-\beta_\rme H^{(n)}_\rme}]}
\eeq
with $\beta_{\rme}={1}/(k_B T_{\rme})$ the inverse temperature of the $\rme=\rmc, \rmh$ environment (here $k_B$ is the Boltzmann constant). We have also assumed the hierarchy of temperatures $T_{\rmc}< T_{\rmh}$. The working medium is assumed to be initialized in a thermal state at initial temperature $T_{\rms}$ such that $T_{\rmc} < T_{\rms} < T_{\rmh}$. With reference to Figure~\ref{fig1}, our Otto cycle is implemented with the following steps:
\begin{description}
	\item[ Stroke 1--Compression] We let the initial internal energy of the working medium be $E_0$. The oscillator frequency is changed from $\omega_{\rmc}$ to $\omega_{\rmh}$ in isolation from any environment. The final energy is $E_1$ and the work \emph{done on} the medium is $W_{\rmin}=E_0-E_1 <0$.
	\item[Stroke 2a--Contact with hot environment] The engine interacts with a hot-environment particle and the final internal energy is $E_2$. The engine absorbs the heat $Q_{\rmin}=E_2-E_1 >0$.
	\item[Stroke 2b--Intra-environment interaction] The intra-environment interactions may propagate some memory of the medium's state across the environment, and feed it back at a later stage. This step has no direct effect on the thermodynamics of the engine.
	\item[Stroke 3--Expansion] The frequency of the oscillator is changed from $\omega_{\rmh}$ back to $\omega_{\rmc}$ in isolation from any environment. The final energy is $E_3$ and the work \emph{performed by} the engine is $W_{\rmout}=E_2-E_3>0$.
	\item[Stroke 4a--Contact with cold environment] The engine interacts with a cold-environment particle and the final internal energy is $E_4$. The engine has transferred an amount of heat $Q_{\rmout}=E_4-E_3 <0$ to the environment.
	\item[Stroke 4b--Intra-environment interaction] This stroke is similar to stroke {\bf 2b}.
\end{description}
The final state of the medium becomes the initial state of a new cycle and the steps are iterated, involving new environmental particles. The dynamics thus proceeds through discrete time steps, each of them being a full iteration of the Otto cycle. At the end of each cycle, we compute the {\it power} output of a cycle, and its {\it efficiency}. By denoting with $\ct$ the total duration of one cycle, the power output is $P=(W_{\rmin}+W_{\rmout})/\ct$, while the efficiency reads $\eta=(W_{\rmin}+W_{\rmout})/Q_{\rmin}$. We ignore any decoherence channel affecting the oscillator or the spins by claiming that the overall evolution takes place in a time that is shorter than the smallest time-scale set by such mechanisms. 

Let us define as $n_k$ ($k=0,\dots,4$) the average occupation number at the beginning ($k=0$) and after step $k \geq 1$ of the protocol, such that $E_k=\hbar \omega_{\rme} (1/2 + n_k)$, with $\omega_{\rme}=\omega_{\rmc}$ [$\omega_{\rme}=\omega_{\rmh}$] at the beginning and after strokes 3 and 4 [1 and 2]. 
Using (\ref{eq:IntEn})-(\ref{eq:heat}), we have
\begin{equation}
\eta = \frac{E_2 - E_3 + E_0-E_1}{E_2-E_1}=1-\frac{\omega_{\rmc}(n_3-n_0)}{\omega_{\rmh}(n_2-n_1)}.
\end{equation}
If the work transformations are performed adiabatically, the populations remain unchanged and $n_1=n_0$ and $n_2=n_3$. The theoretical efficiency thus reads
$\eta_{\rmth}=1-{\omega_{\rmc}}/{\omega_{\rmh}}$,
irrespectively of the details of the heat exchanges.

\subsection{Work transformations}
\label{sec:work}

The work strokes are implemented through a unitary transformation on the engine alone, isolated from the cold or hot environment. A theoretical description of such processes was developed in \cite{Husimi53} and further extended in \cite{Deffner08}. In the following, we summarise the key steps of such approaches, which represent the basis for our implementation of the work strokes. 

We wish to find a wave-function $\psi(x,t)$ satisfying the Schr\"odinger equation
\beq
\label{eq:schro}
i \hbar \partial_t \psi(x,t) = {H}_{\rms}(t) \psi(x,t)
\eeq
within the time interval $[0,\tau]$, with $\omega(0)=\omega_1$ and $\omega(\tau)=\omega_2$. In the following, $\omega_1$ and $\omega_2$ will be either $\omega_{\rmc}$ or $\omega_{\rmh}$ depending on which work transformation is being performed. 
The Hamiltonian in (\ref{eq:HO_Hamiltonian}) can be written, at any \emph{fixed} time $t$, as 
\beq
\label{eq:HO_Hamiltonian2}
{H}_{\rms}(t)=\hbar \omega(t) \big( 1/2 + {a}^{\dag}(t) {a}(t) \big),
\eeq  
where the operators
\begin{equation}
{a}(t)= \sqrt{\frac{m \omega (t)}{2 \hbar}} {x} + i \sqrt{\frac{1}{2 m \hbar \omega (t) }} {p} 
\end{equation}
and $a^\dag (t)=[a(t)]^\dag$
depend explicitly on time.
From (\ref{eq:HO_Hamiltonian2}), we obtain the instantaneous eigenvalues $E^t_n = \hbar \omega(t) (1/2 + n(t))$ and the wave-function $\phi_n^t(x)$ of its eigenvectors, which are just a slight generalization of the solutions for the time-independent quantum harmonic oscillator. Explicitly
\beq
\label{eq:HOeigenstates}
\phi_n^t(x) = \sqrt[4]{\frac{m \omega(t)}{\pi \hbar}} \frac{1}{\sqrt{2^n n!}} e^{- \frac{m \omega(t)}{2 \hbar} x^2} H_n \bigg(x \sqrt{\frac{m \omega(t)}{\hbar}} \bigg),
\eeq
where $H_n(z)$ is the $n^\mathrm{th}$ Hermite polynomial of argument $z$. The superscript $t$ aims at reminding that $t$ here plays just the role of a label. 
(\ref{eq:schro}) admits solutions satisfying the Gaussian ansatz
\begin{equation}
\psi(x,t) = \exp \big[ i\big( A x^2 + 2 B x +C \big)/2\hbar \big],
\end{equation}
where the time-dependence is hidden in the coefficients $A(t),\, B(t),\, C(t)$.
By inserting this formula into (\ref{eq:schro}), we obtain the system of differential equations 
\bearr
\frac{d A}{d t} &= - \frac{A^2}{m} - m\omega^2(t),
\label{eq:a}
\\
\label{eq:b}
\frac{d B}{d t} &= -\frac{A}{m}B,
\\
\label{eq:c}
\frac{d C}{d t} &= i \hbar \frac{A}{m} - \frac{1}{m} B^2.
\eearr
Equation (\ref{eq:a}) can be mapped into the equation of motion of a classical time-dependent oscillator with amplitude $X(t)$, through the substitution $A=m \dot{X}/X$. Explicitly
\beq
\label{eq:X}
\frac{d^2 X}{dt^2} + \omega^2(t) X = 0.
\eeq 
Once a parameterization is chosen for $\omega(t)$, all the unknown coefficients can be found by direct integration. In \cite{Deffner08} it is shown that by choosing the parameterization 
\beq
\omega^2(t) = \omega_2^2 +{t}(\omega^2_1-\omega^2_2)/{\tau},
\eeq
an analytic solution to such problem can be found. We refer to the mentioned reference for the full expression. Another key result is the expression of the {propagator}~\cite{Husimi53}
\beal
\label{eq:Uprop}
U(x,\tau |x_0,0)=
\sqrt{\frac{m}{2 \pi i \hbar X(\tau)}}& \exp \bigg[ \frac{i m}{2 \hbar X(\tau)} (\dot{X}(\tau) x^2 - 2 x x_0 + Y(\tau) x_0^2) \bigg],
\eeal
where now $X(t)$ and $Y(t)$ are two specific solutions of (\ref{eq:X}) satisfying the boundary conditions 
\beal
&X(0)=0, \qquad \dot{X}(0)=1,
\\
&Y(0)=1, \qquad \dot{Y}(0)=0.
\eeal
With the propagator $U(x,\tau | x_0, 0)$, we now have all the tools to describe the effect of the work transformation $\omega_{1} \to \omega_{2}$ (for arbitrary values of $\omega_{1,2}$) on the medium's state $\rho(x,y;t)$. We have
\beal
\label{eq:work_trans}
&\rho(x_0,y_0;0) \mapsto 
\\
&\rho(x,y;\tau)=
 \int U(x,\tau | x_0,0) \rho(x_0,y_0;0) U^{\dag}(y,\tau | y_0,0) \rmd x_0 \rmd y_0  
\eeal

One further step is required, with the aim of making the above transformation amenable to numerical treatment, namely the expansion of both the density matrix $\rho$ and the propagator on the basis given by the eigenfunctions in (\ref{eq:HOeigenstates}). Let us define
\bearr
\rho_{mn}(t) = \langle \phi_m^t | \rho(t) | \phi_n^t \rangle,
\\
\label{eq:Umn}
U_{mn} = \langle \phi_m^{\tau} | U(\tau,0) | \phi_n^0 \rangle,
\eearr
where we omitted the position dependencies since they are integrated over in the scalar products. It should be stressed that the $U_{mn}$ elements are computed by taking scalar products with two \emph{different} sets of eigenfunctions, the effect of $U(\tau,0)$ being precisely that of implemen\-ting the transformation from one Hamiltonian to another. Equation (\ref{eq:work_trans}) then becomes
\beq
\rho_{mn}(0) \mapsto \rho_{kl}(\tau) = \sum_{mn} U_{km} \rho_{mn}(0) U^{\dag}_{nl}.
\eeq

\begin{figure}[t]
\begin{center}
\includegraphics[width=0.65\columnwidth]{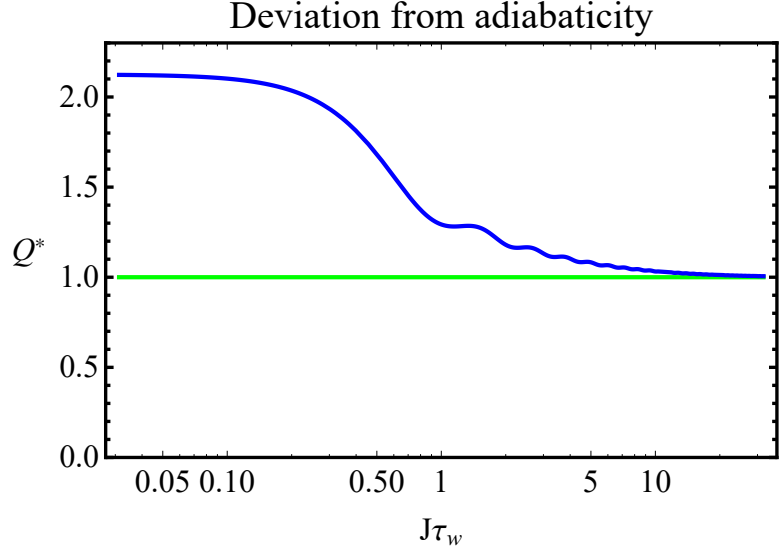}
\end{center}
\caption{Deviation from the adiabatic regime, as captured by the $Q^*$ factor (\ref{eq:Qstar}), as a function of the work stroke duration $\tau$, for $\omega_1=1$ and $\omega_2=4$. The straight green line represents the limit $Q^*=1$ for $\tau \to \infty$. \label{fig2}}
\end{figure}

The transition probabilities from the initial to the final eigenstates are readily obtained as $P^{\tau}_{m,n}=|U_{mn} |^2$. In \cite{Husimi53} an expression for their generating function
\beq
P(u,v) = \sum_{m,n} u^m v^n P^{\tau}_{m,n},
\eeq
such that  
\beq
P^{\tau}_{m,n} = \frac{1}{m! n!} \frac{\partial^{m+n} P(u,v)}{\partial u^m \partial v^n} \bigg{|}_{u=v=0},
\eeq
is provided as
\beq
P(u,v) = 
\sqrt{\frac{2}{Q^* (1-u^2)(1-v^2) + (1+u^2)(1+v^2)-4uv}}.
\eeq
Remarkably, the above expression depends on the details of the parametrization $\omega(t)$ only through the factor $Q^*$, whose expression for the most general transformation is~\cite{Deffner08}
\beq
\label{eq:Qstar}
Q^*=\frac{ \omega_1^2 \big( \omega_2^2 X(\tau)^2 + \dot{X}(\tau)^2 \big) + \big( \omega_2^2 Y(\tau)^2 + \dot{Y}(\tau)^2 \big) }{2 \omega_1 \omega_2}.
\eeq
We have $Q^* \to 1$ ($Q^*$ increasingly greater than 1) for $\tau \to \infty$ ($\tau\to 0$), as shown in Figure~\ref{fig2}.

The following special cases are of particular interest:
\begin{itemize}
	\item \emph{No transformation is performed}, $\omega_2 = \omega_1$. It can be shown that the propagator in (\ref{eq:Uprop}) becomes the identity operator and thus $U(x,\tau | x_0,0) = \delta(x-x_0)$. The matrix elements in (\ref{eq:Umn}) are $U_{mn}=\delta_{mn}$, as the initial and final eigenbases coincide.
	\item \emph{Sudden quench}, $\tau \to 0$. Also in this case $U(x,\tau | x_0,0) \to \delta(x-x_0)$, because the transformation is so quick that the density matrix is left unchanged. Its matrix elements $\rho_{mn}$, however, undergo a unitary change of basis through the matrix $U_{mn} =  \langle \phi_m^{(2)} | \phi_n^{(1)} \rangle$, where the superscripts refer to the frequencies $\omega_1, \,\omega_2$.     
	\item \emph{Adiabatic transformation}, $\tau \to \infty$. The initial eigenstates are mapped one-to-one to the final ones, infinitely slowly, up to a phase factor. The propagator becomes $U(\tau,0)= \sum_n e^{i \varphi_n}\ket{\phi^{\tau}_n} \bra{\phi^0_n} $, and thus $|U_{mn}|^2 = \delta_{mn}$.
\end{itemize}

From now on, we will denote the duration $\tau$ of the work transformations by $\tau_{\rmw}$.

\subsection{Heat exchanges: collisional model}
\label{sec:heat_coll}

Let us now introduce the medium-environment and intra-environment interactions, which are implemented through a \emph{collisional model.} We assume that each medium-environment event takes place through the unitary interaction of the oscillator with a single environmental particle at a time. This is what we refer to as a \emph{collision}. We also assume that the working medium never interacts twice with the same environmental particle: after each collision, the medium interacts with a fresh environmental particle. The unitary $V_{\rms \rme }= e^{-\frac{i}{\hbar} H_{\rms \rme} \tau_{\rmse}}~(\rme = \rmc,\rmh)$ through which the interaction takes place is generated by the resonant excitation-conserving Hamiltonian
\begin{equation}
\label{eq:seint}
H_{\rms \rme } = J \big(a \sigma_{\rme }^+ + a^{\dag} \sigma_{\rme }^- \big), 
\end{equation}
where $J$ is the coupling constant and $\tau_{\rmse}$ the interaction time. These parameters are assumed to be the same for both the cold and hot environment.

As mentioned in Sec.~\ref{sec:cycle}, when a collision occurs, the frequency of the working medium matches exactly that of the environmental particle it is interacting with.   
In the most basic, memoryless implementation of such a model, only one particle per environment is retained at any time. Indicating by $\ch_{\rms}$, $\ch_{\rmc}$ and $\ch_{\rmh}$ the Hilbert spaces of the working medium, a cold and a hot particle respectively, the  total Hilbert space is $\ch = \ch_{\rmc} \otimes \ch_{\rms} \otimes \ch_{\rmh}$.  With reference to Figure~\ref{fig1} {\bf (a)}, suppose the working medium is in state $\rho_{\rms}$ at the beginning of iteration $n$ of the cycle, and interacts with the $n^\mathrm{th}$ cold particle initially in state $\rho_{\rmc}^{(n)}$ according to the scheme
\beq
\rho_{\rmc}^{(n)} \otimes \rho_{\rms} \otimes \rho_{\rmh}^{(n)} \to 
\tilde{\rho}_{\rmcsh} = (V_{\rmsc} \otimes \id_{\rmh}) \big( \rho_{\rmc}^{(n)} \otimes \rho_{\rms} \otimes \rho_{\rmh}^{(n)} \big) 
(V_{\rmsc}^{\dag} \otimes \id_{\rmh}),
\eeq
where $\id_{\rmh}$ is the identity matrix in the hot particle's Hilbert space~\cite{remark}.
After the interaction, we take the reduced states $\tilde{\rho}_{\rms } = \Tr_{\rmc, \rmh}[\tilde{\rho}_{\rmcsh}]$ and $\tilde{\rho}_{\rmc}^{(n)} = \Tr_{\rms, \rmh}[\tilde{\rho}_{\rmcsh}]$ and use them to compute the thermodynamic quantities introduced in Sec.~\ref{sec:cycle}. Particle $\rho_{\rmc}^{(n)}$  is then discarded and a new one $\rho_{\rmc}^{(n+1)}$ is included in the model in its place.

We now take a step further and introduce intra-environment collisions, thus allowing the environments to carry over memory of past interactions with the medium, and thus allowing for possible non-Markovian effects to take place. We thus wish to consider two particles per environment, at any given time. In order to do so, we need to extend the Hilbert space we work with to $\ch = \ch_{\rmc,b} \otimes \ch_{\rmc,a} \otimes \ch_{\rms} \otimes \ch_{\rmh,a} \otimes \ch_{\rmh,b}$, where the additional subscript $a$ stands for the first (hot or cold) environmental particle interacting with the engine, and $b$ stands for the second one, that is particles $n$ and $n+1$ in our example. Before we trace it away, the $n^\mathrm{th}$ environmental particle undergoes a further collision with particle $n+1$. Such collision occurs according to the propagator $V_{\rmee} = e^{-\frac{i}{\hbar} H_{\rme \rme} \tau_{\rmee}}$ with $H_{\rmee}$ the Heisenberg Hamiltonian
\begin{equation}
\label{eq:eeint}
H_{\rmee} = J_{\rmee} \big(\sigma_n^{\rmx} \sigma_{n+1}^{\rmx} + \sigma_n^{\rmy} \sigma_{n+1}^{\rmy} + \sigma_n^{\rmz} \sigma_{n+1}^{\rmz}\big), 
\qquad (\rmee = \rmcc,\rmhh).
\end{equation}   
We have introduced the coupling constant $J_{\rmcc}$ $(J_{\rmhh})$ and interaction time $\tau_{\rmcc}$ $(\tau_{\rmhh})$ for the cold (hot) environment. As discussed in Refs.~\cite{Scarani02, McCloskey14, Pezzutto16}, the interaction acts effectively as a \emph{partial swap}, exchanging the states of the two particles with probability $\sin^2(2 J_{\rmee} \tau_{\rmee})$. In particular, a perfect swap is achieved for $J_{\rmee} \tau_{\rmee}=\pi/4$.

Continuing with our example, after the application of $V_{\rms \rmc}$ and $V_{\rmcc}$, the working medium and $(n+1)^\mathrm{th}$ environmental particle will be, in general, in a {correlated state}, which we dub $\tilde{\rho}^{(n+1)}_{\rms \rmc}$. This occurs {even if they did not interact directly yet.} After tracing away the (cold) $n^\mathrm{th}$ environmental particle, shifting particle $n+1$ from position $(c,b)$ to $(c,a)$ in the Hilbert space, and including a new particle -- the $(n+2)^\mathrm{th}$ -- at position $(c,b)$, the global state can be written as
$\rho_{\rmc}^{(n+2)} \otimes \tilde{\rho}^{(n+1)}_{\rms \rmc} \otimes \rho_{\rmh}^{(n)}
\otimes \rho_{\rmh}^{(n+1)}$. 

This completes the description of one full heat stroke. The device is now ready for the next stroke, which will be a work one. The interactions between the working medium and the hot environment, and between particles pertaining to the hot environment itself, would occur in exactly the same way. Therefore, at the end of a full cycle, composed of all the steps of Sec.~\ref{sec:cycle}, the global state reads $\rho_{\rmc}^{(n+2)} \otimes \tilde{\rho}^{(n+1)}_{\rms \rmc \rmh} 
\otimes \rho_{\rmh}^{(n+2)}$. 
More details on this model of system-environment interaction can be found in \cite{Pezzutto16}.

Finally, the total cycle duration is $\ct=2(\tau_{\rmw}+\tau_{\rmse})$, taking into account only the steps in which the engine is directly involved and assuming the intra-environment interactions to occur at the same time as the work strokes.

\section{Results}
\label{sec:results}

We present here the results on the engine performance and the possible influence of non-Markovianity on its operations. First, we study the degree of non-Markovianity ensuing from the engine dynamics and its dependence on intra-environment interactions. We then investigate the crossover from adiabatic to sudden work strokes in the purely Markovian regime, focusing on issues of irreversibility. Finally we address the performance of the engine, highlighting an interesting transition from a thermal machine to a refrigerator. 

In what follows, unless otherwise stated, we use units such that $\hbar=k_B = 1$, and take $J=J_{\rmcc}=J_{\rmhh}=1$, which we can do without affecting the generality of our results. The temperatures of the environments are $T_{\rmc}=0.1$ and $T_{\rmh}=10$, giving a Carnot efficiency of 0.99 and a Curzon-Ahlborn efficiency of 0.9 as theoretical upper bounds. The engine is initialized in a thermal state at $T_{\rms}=0.5$ unless otherwise stated. While the choice of initial temperature is only marginally relevant, the initial absence of coherence in the energy eigenbasis impacts significantly the behaviour of the engine.  

We chose a moderate interaction strength between the working medium and the environments ($J\tau_{\rmse}=0.3$), so that the heat exchanged per cycle remains small yet non negligible compared to the work being performed. The values of the environmental frequencies are $\omega_{\rmc}=1$ and $\omega_{\rmh}=4$, which are such that the work being performed is significant and the adiabatic regime ($\tau_{\rmw} \to +\infty$) is approximated well at $\tau_{\rmw}=16$ and very well at $\tau_{\rmw}=32$. The gap between $\omega_{\rmh}$ and $\omega_{\rmc}$ is nontheless big enough that, in the sudden quench regime, the $Q^*$ factor is appreciably different from 1 (in fact surpassing 2, as it can be seen from Figure~\ref{fig2}). The theoretical efficiency in the adiabatic case is thus $\eta_{\rmth}=0.75$. In what follows, we choose the the eigenbasis $\{\ket{0},\ket{1} \}$ of the Hamiltonian $\hbar \omega_{\rme}  \sigma^{\rmz}_{\rme}/2$ to represent the states of the environments.

As the initial temperature $T_{\rms}$ is low, the initial populations decay quite fast, becoming negligible (below machine precision) above the $20^\mathrm{th}$ energy level of the oscillator. Therefore, in most of the simulations we could safely truncate the computational space at level $30$, checking that the matrices representing the unitaries $U, \, V_{\rmse}$, and $V_{\rmee}$ in the truncated space remain approximately unitary, and all states have unit trace. We performed tests extending the Fock space up to level $50$ to confirm that the results that we report here were not appreciably different than those obtained using the stated computational space.

\subsection{Non-Markovianity of the engine dynamics}

\begin{figure}[t]
\includegraphics[width=1.\columnwidth]{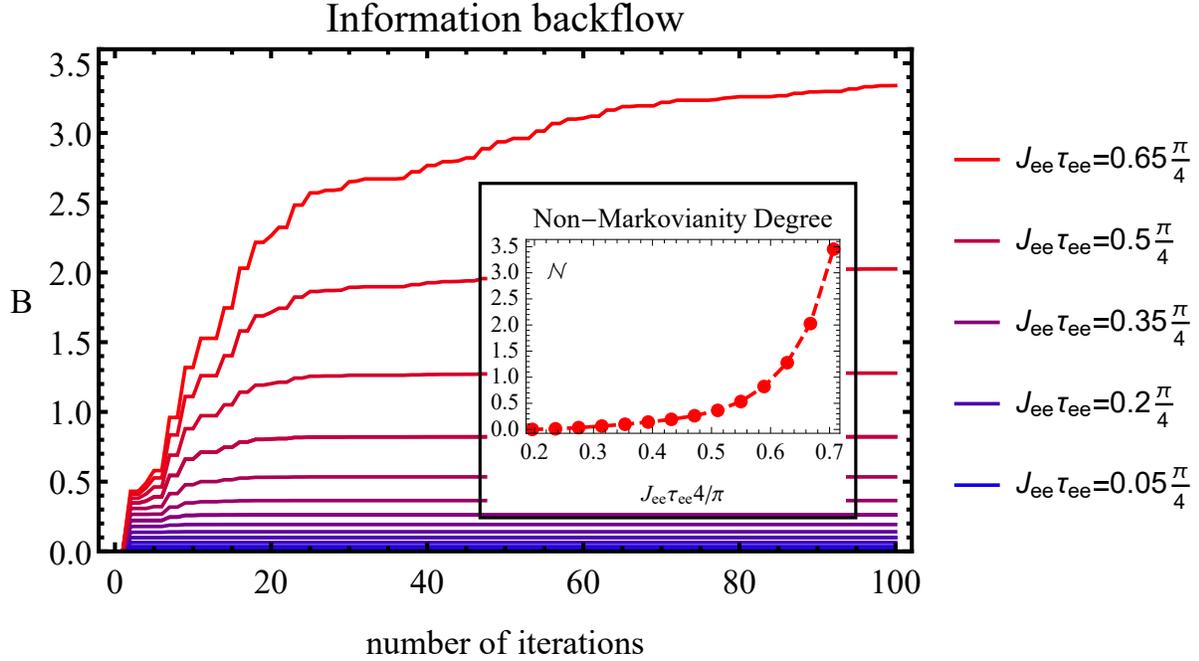}~~~
\caption{Information backflow $B(t)$ (\ref{eq:infobackflow}) capturing the time evolution of the degree of non-Markovianity $\cn$ (\ref{eq:N}), with intra-environment interaction $J_{\rmee} \tau_{\rmee}$ increasing from blue to red (bottom to top). Nearly adiabatic work strokes ($J\tau_{\rmw}=32$). The pair of pure initial states $\ket{\psi^{\pm}_{\mathrm{test}}}$ (\ref{eq:psitest}) was used to effectively detect non-Markovianity. 
Inset: final $\cn$ against the intra-environment interaction $J_{\rmee} \tau_{\rmee}$. The dashed line is a guide for the eyes.}
\label{fig3}
\end{figure}

Recently, the issue of non-Markovianity of quantum dynamics has received considerable attention aimed at characterizing the phenomenology of non-Markovian open-system dynamics through general tools of broad applicability. Such efforts are based on the formal assessment of the various facets with which non-Markovianity is manifested.

One of such approaches, introduced in Refs.~\cite{Breuer09,Laine10}, is based on the concept of {information backflow.} 
Let us introduce the {trace distance} between two states~\cite{NielsenChuang10} 
\beq
D(\rho_1,\rho_2) := \half \| \rho_1 - \rho_2 \| \,,
\eeq  
where $\| A \| = \Tr \sqrt{A^{\dag} A}$ is the trace-1 norm of operator $A$, and $\rho_{1,2}$ are two density matrices of the system under scrutiny. The trace distance is a metric in the space of density matrices, closely related to their {distinguishability}: a value of $D(\rho_1,\rho_2)=1$ implies perfect distinguishability. 

Any completely positive trace-preserving (CPTP) map is a {contraction} for the {trace distance}. This is the key idea for the quantification of non-Markovianity based on {information backflow}: Markovian maps cannot increase the distinguishability of any two given states. If, however, one can find a pair of initial states and a time $t$ for which contractivity is violated, thus resulting in 
\beq
\label{eq:markovian}
\sigma(t)=\frac{d D(\rho_1(t),\rho_2(t)) }{dt} > 0,
\eeq
this is held as a signature of non-Markovianity in the dynamics. Such criterion can be used to build a quantitative measure as~\cite{Breuer09}, the \emph{degree of non-Markovianity}
\beq
\label{eq:N}
\cn:= \max_{ {\{ } \rho_1,\rho_2 {\}} } \int_{\Sigma_+}  \sigma(t) \rmd t,
\eeq
where $\Sigma_+$ is the time window where $\sigma(t)>0$, and we should maximize over the choice of initial states. To observe how non-Markovianity appears during the time-evolution, a useful quantity is the total backflow of information from time $t_0$ up to time $t$
\beq
\label{eq:infobackflow}
B(t):= \max_{ {\{ } \rho_1,\rho_2 {\}} } \int_{\Sigma_+,t'=-\infty}^{t`=t_0}  \sigma(t') \rmd t',
\eeq
closely related to the degree of non-Markovianity since $\cn = B(+\infty)$.

\begin{figure}[t]
\includegraphics[width=1.\columnwidth]{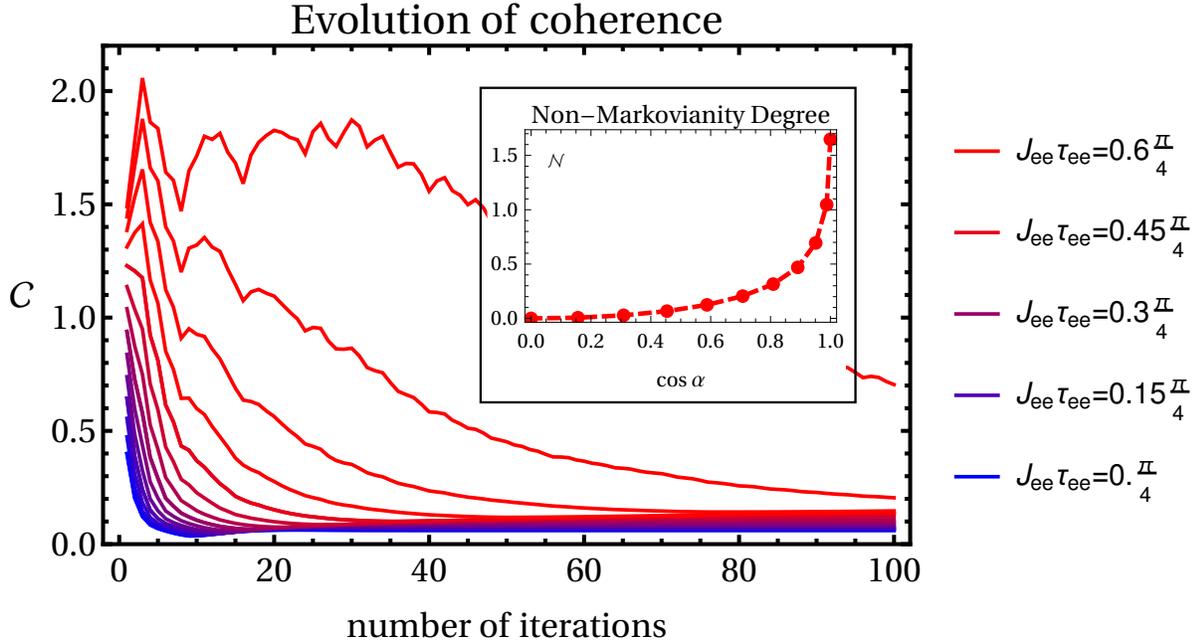}
\caption{Time evolution of the coherence $\cc$ in the density matrix of the working medium (\ref{eq:coherence}), with intra-environment interaction $J_{\rmee} \tau_{\rmee}$ increasing from bottom to top curve. We have taken a pure initial states $\ket{\psi^{\pm}_{\mathrm{test}}}$. The work strokes are nearly adiabatic owing to the choice $J\tau_{\rmw}=32$. Inset: final $\cn$ against the coherence $\cos\alpha$ in the pair of initial states $\ket{\psi_{\alpha}},\ket{\psi^{\perp}_{\alpha}}$, defined in (\ref{eq:psialpha}), used to detect non-Markovianity. The intra-environment interaction is $J_{\rmee} \tau_{\rmee}=0.65 \pi/4$. The dashed line is a guide for the eyes.}
\label{fig4}
\end{figure}

While finding the optimal pair of initial states is in general  challenging, the task is often simplified owing to the result reported in Ref.~\cite{Wissmann12}, where it is proven that the optimal states must be orthogonal and belonging to the boundary of the state space. In our case, however, the state of the engine is represented by a very large Hermitian matrix and the maximization is an extremely demanding task. We thus heuristically choose a pair of pure orthogonal states $\ket{\psi^{\pm}_{\mathrm{test}}}$, guided by the analogy with the spin-1/2 particle case in which often the optimal pair is $\ket{\pm} = (\ket{0} \pm \ket{1})/\sqrt{2}$~\cite{McCloskey14,Pezzutto16}. We thus consider
\beq
\label{eq:psitest}
\ket{\psi^{\pm}_{\mathrm{test}}} = \frac{\ket{0} \pm \ket{10}}{\sqrt{2}},
\eeq
as we found that pure states in the form $(\ket{0} \pm \ket{n})/\sqrt{2}$, which have a high degree of coherence in the energy eigenbasis, appear to be effective in the establishment of lower bounds to the non-Markovianity measure, thus providing a valuable insight on the non-Markovian character of the dynamics. Needless to say, such lower bound would quantitatively depend on the actual choice of state $\ket{n}$. However, this is immaterial for our goals, as we only aimed at identifying an instance of initial pair of states for which the contractivity of the trace distance is violated. 

\begin{figure}[b]
\begin{center}
{\bf (a)}\hskip7cm{\bf (b)}\\
\includegraphics[width=0.45\columnwidth]{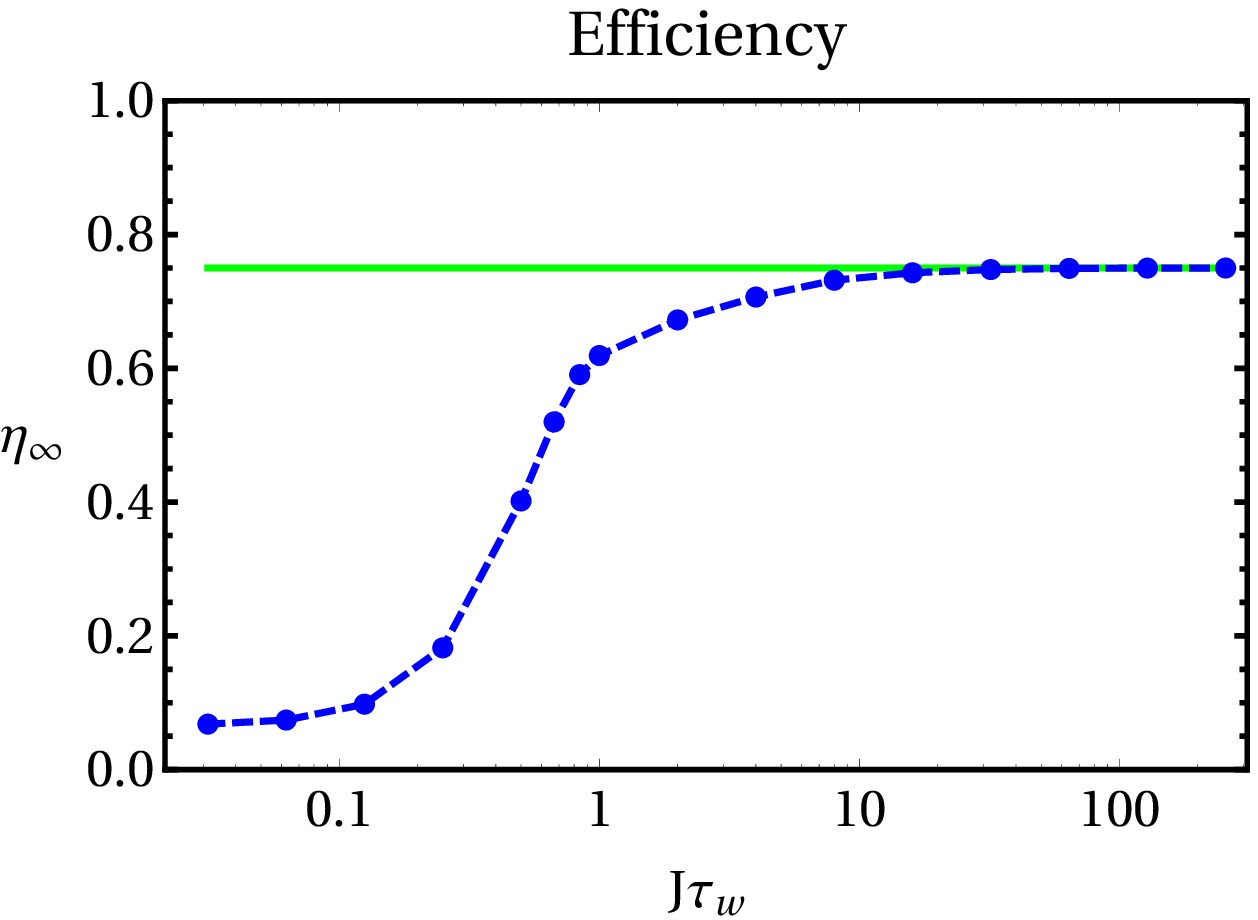}~~~~
\includegraphics[width=0.45\columnwidth]{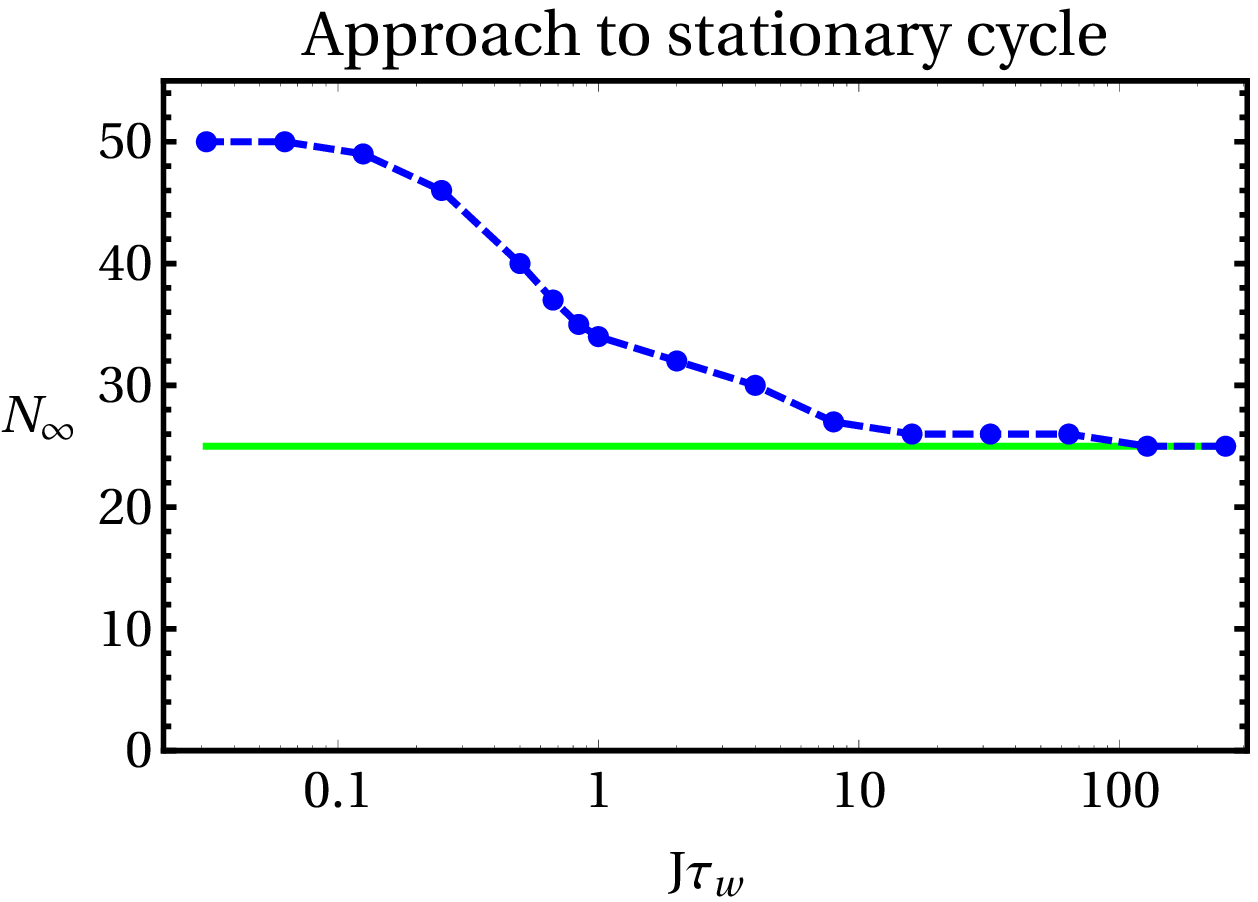}
\end{center}
\caption{{\bf (a)} The blue dots show the stationary cycle efficiency $\eta_{\infty}$ against the dimensionless duration of the work stroke $J\tau_{\rmw}$.  The dashed line is a guide for the eyes, the solid green line represents the theoretical adiabatic efficiency $\eta_{\mathrm{th}}=1-\omega_\rmc/\omega_{\rmh}$.
{\bf (b)} We show the dependence of the number of iterations $N_{\infty}$ required to reach the stationary cycle on $J\tau_{\rmw}$. The dashed line is a guide for the eyes, the solid green  line shows the value for $\tau_\rmw \to \infty$.
}
\label{fig5}
\end{figure}

Figure~\ref{fig3} presents the behaviour of $\cn$ against the intra-environment interaction strength and time in the case of adiabatic work strokes. The non-Markovian behaviour is intrinsically a property of the dynamics during the transient to stationary state. Figure~\ref{fig4} shows the dynamics of the total internal coherence of the engine, quantified by~\cite{Baumgratz}
\beq
\label{eq:coherence}
\cc := \sum_{i \neq j} |\rho_{ij}|.
\eeq

The coherence in the stationary state settles to a quite small value, irrespective of the initial state. Furthermore, the more non-Markovian the dynamics, the longer coherences survive. This is most likely a direct consequence of the fact that the interaction with environments inducing non-Markovian dynamics slows down the approach to the stationary state (see also Figure~\ref{fig7}). The inset of Figure~\ref{fig4} shows the relation between non-Markovianity and the initial coherence present in the engine, when initialized in  states
\beq
\label{eq:psialpha}
\ket{\psi_{\alpha}}=\cos \alpha \ket{0} + \sin \alpha \ket{10}
\eeq
and $\ket{\psi^{\perp}_{\alpha}}$ orthogonal to $\ket{\psi_{\alpha}}$, with $\alpha = \pi /4 \times 0.1 m$ ($m=0,1, \dots,10$). Note that the pair of states $\ket{\psi^{\pm}_{\mathrm{test}}}$ (\ref{eq:psitest}) is obtained for $\alpha = \pi/4$. The connection between the presence of coherence in the initial states and their effectiveness in the revelation of non-Markovianity is very strong.

\subsection{Performance of the engine}

\begin{figure}[t]
\begin{center}
\includegraphics[width=0.65\columnwidth]{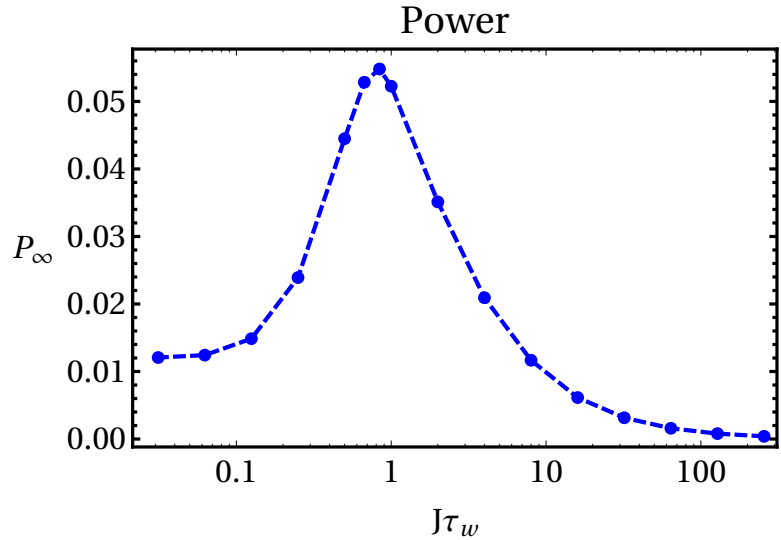}
\end{center}
\caption{Power output against the dimensionless duration of the work stroke $J\tau_{\rmw}$. The dashed line is a guide for the eyes. The power vanishes for $\tau_{\rmw}\to \infty$, as the efficiency approaches the limit $\eta_{\rmth}$ while the cycle duration grows as $\sim 2 \tau_{\rmw}$. In the sudden quench limit, instead, it approaches a finite value, being $\eta$ non-zero for $\tau_{\rmw} \to 0$ while the cycle duration is $\sim 2 \tau_{\rmse}$.  \label{fig6}}
\end{figure}

Figures~\ref{fig5} and~\ref{fig6} summarize the behaviour of the engine in the Markovian regime, with no intra-environment interactions, focusing on the crossover from adiabatic to sudden quench work strokes. A general feature we always observe is that the dynamics of always ends up in a stationary cycle: after a certain number of iterations, the density matrix of the engine keeps cycling through the same four states repeatedly and indefinitely, as it goes through the Otto cycle. The stationary state depends on the parameters of the model (frequencies and temperatures of the environments) and is independent on the initial engine state, as well as on the system-environment coupling, which only affects the pace at which the stationary cycle is reached. We can see that the stationary cycle efficiency $\eta_{\infty}$ reaches the expected limit $\eta_{\rmth}$ in the adiabatic case, and decreases as we depart from adiabaticity. The duration of the work strokes $\tau_{\rmw}$ also affects the number of iterations $N_{\infty}$ it takes for the engine to reach the stationary regime, which grows as we approach the sudden quench regime. This further indicates a drop of the engine performance as we move away from adiabaticity. The power output per single iteration $P_{\infty}$, however, has a maximum around $\tau_{\rmw}=1$, since at that point the efficiency deviates only slightly from $\eta_{\rmth}$.  

\begin{figure}[t]
\begin{center}
\includegraphics[width=0.7\columnwidth]{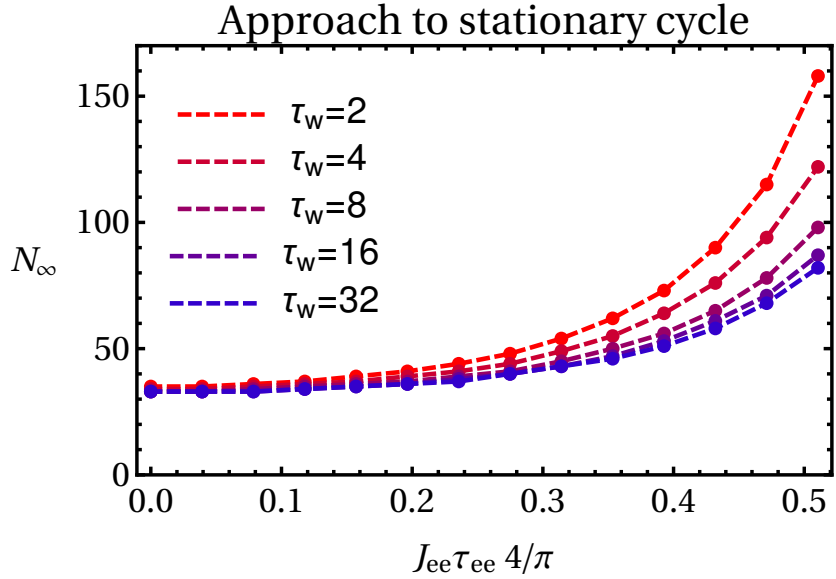}
\end{center}
\caption{Number of iterations $N_{\infty}$ needed to reach stationarity against the dimensionless intra-environment interaction time $J_{\rmee} \tau_{\rmee}$ and for growing values (in units of the coupling strength) of the duration $\tau_\rmw$ of the work strokes. The dashed lines are guides to the eyes.
\label{fig7}}
\end{figure}

\begin{figure}[t]
\begin{center}
{\bf (a)}\hskip7cm{\bf (b)}\\
\includegraphics[width=0.45\columnwidth]{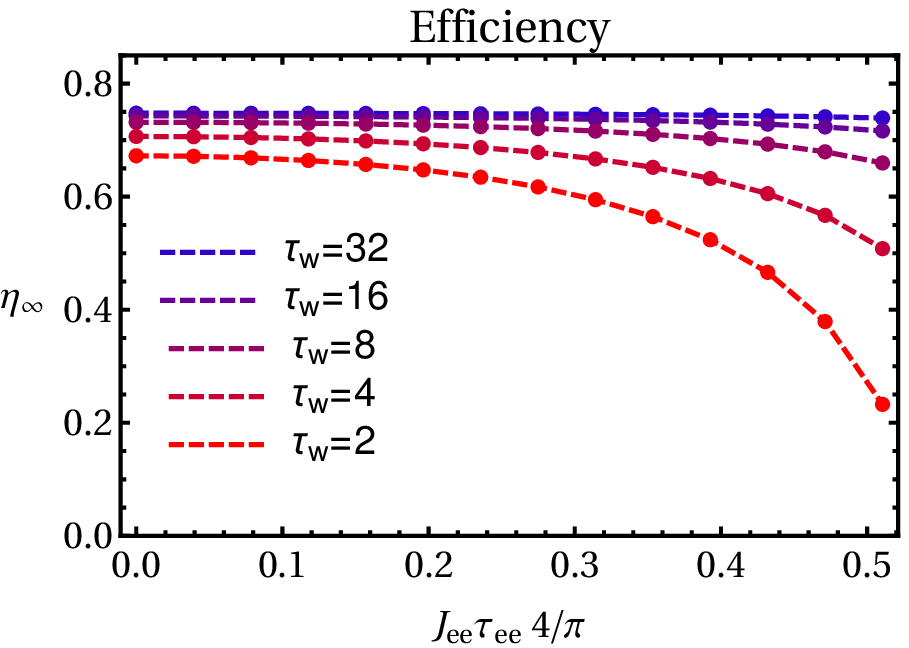}~~~~
\includegraphics[width=0.45\columnwidth]{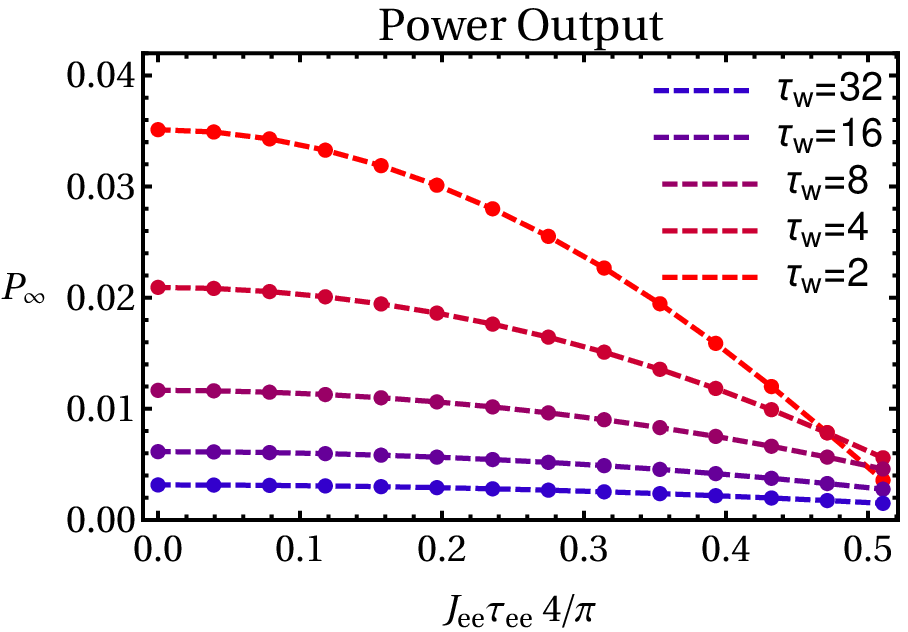}
\end{center}
\caption{Stationary cycle efficiency $\eta_{\infty}$ {\bf (a)} and power output {\bf (b)} against the intra-environment interaction $J_{\rmee} \tau_{\rmee}$. The dashed lines are guides for the eyes. The duration $\tau_{\rmw}$ of the work strokes decreases from the blue to the red curve (top to bottom).
\label{fig8}}
\end{figure}

Figure~\ref{fig7} and~\ref{fig8} present the behaviour of the performance in the most general case of the engine operating with non-adiabatic work strokes and non-Markovian environments. Non-Markovianity seems to always affect negatively the performance, but it does so more pronouncedly as we deviate from the adiabatic regime. In particular, the efficiency in the adiabatic case is mostly independent of the non-Markovian character of the dynamics, approaching in fact $\eta_{\rmth}$, while for smaller durations of the work strokes it drops more neatly as the intra-environment interactions become stronger. The power output, therefore, decreases accordingly. The relation between the phenomenology illustrated here and the interplay between coherence and non-Markovianity deserves a quantitative assessment that goes beyond the scopes of this work.  

\subsection{Characterization of irreversibility}

\begin{figure}[t]
\begin{center}
\includegraphics[width=0.7\columnwidth]{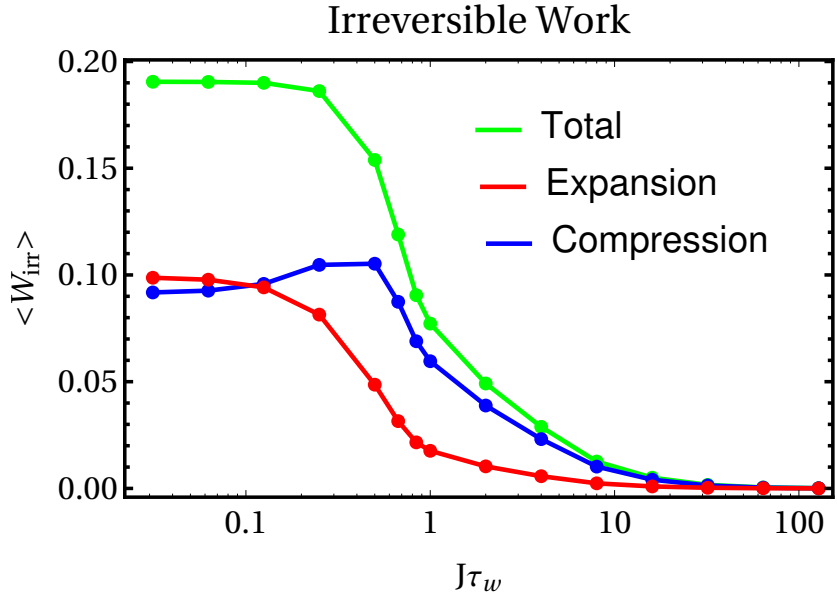}
\end{center}
\caption{Irreversible work $\langle W_\mathrm{irr}\rangle$ against the stroke duration $\tau_\mathrm{w}$ (in units of $J$) [cf. (\ref{eq:Wirr})]. We consider the contribution coming from the compression and expansion stages, as well as the total irreversible work.
\label{fig9}}
\end{figure}

We now wish to investigate further the implications that the crossover from an adiabatic to a sudden-quench transformation has in the Markovian regime, focusing in particular on issues of thermodynamic irreversibility~\cite{Jarzynsky97,Crooks98,Batalhao18}. At the core of a study on irreversible thermodynamical transformation is the concept of {irreversible entropy production} and the closely related notion of {irreversible work}. The latter is the difference between the actual average work exchanged in a transformation, and the amount that would be exchanged if the process were carried out in a reversible fashion. It is defined as
\beq
\label{eq:Wirr}
\langle W_{\mathrm{irr}}\rangle := -\langle W \rangle - \Delta F = -(\langle W \rangle - \langle W_{\mathrm{rev}}\rangle),  
\eeq
where $ \Delta F $ is the free-energy difference and $\langle W_{\mathrm{rev}}\rangle$ is the average work in the adiabatic limit $\tau_{\rmw} \to \infty$. With these definitions at hand, and bearing in mind our sign-conventions, the irreversible work is positive for any transformation occurring in finite time. In the case of our thermodynamic cycle, this holds equally for both kinds of work strokes: in the compression strokes, a positive irreversible work means that more work then in the adiabatic case has to be performed by the external agent. In the expansion strokes, a positive irreversible work means that the work performed by the engine is less then it could be achieved in the reversible case. We have thus calculated the degree of irreversible work attained in both the expansion and compression strokes, and their sum, as $\tau_{\rmw}$ grows. The results valid for Markovian dynamics are shown in Figure~\ref{fig9}. Notice the closeness of the behaviour of $\langle W_\mathrm{irr}\rangle$ with the behaviour of the $Q^*$ factor, Figure~\ref{fig2}, which is indicative of the crucial role that non-adiabaticity plays in the generation of entropy. Apart from insignificant numerical discrepancies due to the finiteness of the sample used for our numerical simulations , the irreversible work associated with the expansion and compression stages display a similar trend, showing less irreversibility for a more pronounced adiabaitic transformation. Needless to say, the condition $\tau_{\rmw}\to\infty$ corresponds to a perfectly reversible process with no associated entropy production. 

\subsection{Temperature effects: from an engine to a refrigerator}

\begin{figure}[t]
\begin{center}
\includegraphics[width=0.8\columnwidth]{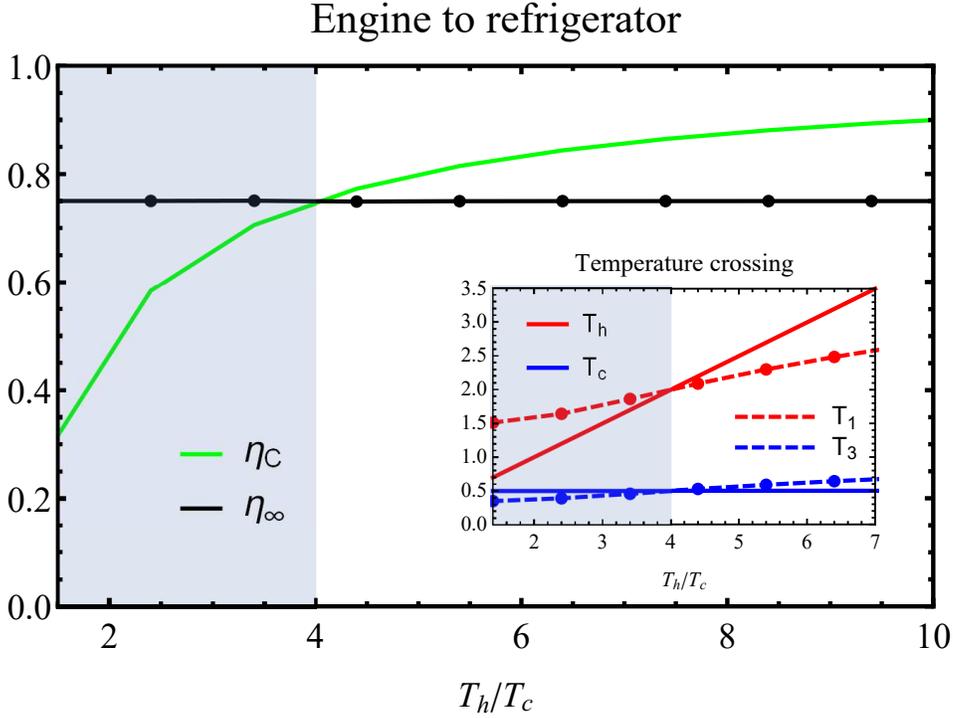}
\end{center}
\caption{Temperature effects in the adiabatic and Markovian regime. As $T_{\rmh}/T_{\rmc}$ drops below $\omega_{\rmh}/\omega_{\rmc}$, the character of the machine changes from engine to refrigerator, the latter being represented by the shaded area. Inset: the transition is explained by the dynamics of the effective temperature of the working medium, in relation with the temperatures of the environments. \label{fig10}}
\end{figure}

All the results presented so far were obtained for a fixed choice of the environmental temperatures. We now explore what happens as we change their respective ratio. A study of the consequences of different choices of this ratio is particularly interesting: as the adiabatic efficiency reads $\eta_{\rmth}=1-\omega_{\rmc}/\omega_{\rmh}$, a choice of parameters such that $\omega_{\rmc}/\omega_{\rmh} < T_{\rmc}/T_{\rmh}$ could result in a better-than-Carnot efficiency (which would be perfectly allowed in light of the non-adiabatic nature of our cycles). This turns out not to be the case, although the ratio $W/Q_{\rmh}$ approaches $\eta_{\rmth}$ in the adiabatic limit for any choice of temperatures. We studied the behaviour of the machine for varying $T_{\rmh}$ -- at a set value of $T_{\rmc}$ -- and frequencies [cf. Figure~\ref{fig10}], finding that if $\omega_{\rmc}/\omega_{\rmh} < T_{\rmc}/T_{\rmh}$, the character of the machine changes from an engine to a refrigerator, as revealed by the switch of the sign of both work and heat flows.

To gain a better understanding of such transition, we studied the evolution of the \emph{effective temperature} of the working medium, defined as the temperature that a quantum harmonic oscillator would have if prepared in a thermal state having the same energy as the working medium of our cycles. 
This leads us to the expression for the effective temperature 
\begin{equation}
T_{\mathrm{eff}} = \omega \bigg[ \ln \Big( \frac{2 E + \omega}{2 E - \omega} \Big) \bigg]^{-1},
\end{equation}
where $E$ is the energy of the working medium. Consider the machine at initial temperature $T_{\rms} \simeq T_{\rmc}$. In the compression stroke work is done on the medium, resulting in an increase of the internal energy and thus of the effective temperature to $T_1$. If $T_{\rmh} > T_1 $, in the ensuing interaction with the hot environment, some heat would flow into the engine, causing the temperature to increase to $T_2 > T_1$. The expansion stroke follows: the engine performs work at the expense of its own internal energy and the effective temperature drops to $T_3$, which is smaller then $T_1$ but still higher then $T_{\rmc}$, which causes heat to be dumped into the cold environment, which completes a cycle.

If, however, the compression stroke results in $T_1 > T_{\rmh}$, during the interaction with the hot environment energy flows from the machine to the environment rather than the other way round. The effective temperature of the medium thus drops to $T_2 < T_1$. Now the expansion stroke occurs, during which the machine loses energy and decreases its temperature to $T_3$. As $T_3$ is most likely smaller then $T_{\rmc}$, during the interaction with the cold reservoir the medium absorbs energy from it, thus completing a refrigeration cycle. 

The transition from engine to refrigerator and the interplay between the various temperatures in the adiabatic case are shown in Figure~\ref{fig10}.

\section{Conclusions}
\label{sec:conc}

In this work we studied the out-of-equilibrium thermodynamics and performance of a quantum Otto cycle employing a harmonic oscillator as working medium The latter is put in interaction with a finite-size environment through a collisional dynamics that may allow for memory effects, and thus for the emergence of non-Markovianity. We explored the crossover from adiabatic to sudden-quench work strokes and found that, while departing from the adiabatic regime induces a drop in the efficiency, it is possible to find an optimal duration of the work strokes such that the power output is maximized. 

The departure from adiabaticity was further characterized through the study of irreversible work. We do not observe better-than-classical performance, at least in the case when both the engine and the environmental particles are initialized in thermal states. Signatures of non-Markovian dynamics are observed in the evolution of the working medium, and even though such memory effects do not impact the performance of the engine at the steady state, they do affect the approach to stationarity, slowing it down. Non-Markovianity is however found to be closely connected with the presence of initial coherences in the energy eigenbasis of the engine.

Finally, by studying the behaviour of the engine across a range of different temperatures, we singled out the parameter regime in which the machine behaves as a refrigerator instead of an engine, and connected this transition with the dynamics of the effective temperature of the working medium.


\ack

MPe thanks the Centre for Theoretical Atomic, Molecular, and Optical Physics, School of Mathematics and Physics, Queen's University Belfast for hospitality during the development of part of this work. We thank Andr\'e Xuereb for helpful discussions. MPe and YO acknowledge support from Funda\c{c}\~{a}o para a Ci\^{e}ncia e a Tecnologia (Portugal) through programmes PTDC/POPH/POCH and projects UID/EEA/50008/2013, IT/QuNet, ProQuNet, partially funded by EU FEDER, from the QuantERA project TheBlinQC, and from the John Templeton Foundation project NQuN (ID 60478).
Furthermore, MPe acknowledges the support from the DP-PMI and FCT (Portugal) through scholarship SFRH/BD/52240/2013.  
MPa acknowledges support from the EU Collaborative project TEQ (grant agreement 766900), the DfE-SFI Investigator  Programme  (grant  15/IA/2864), the  Royal Society Newton Mobility Grant NI160057, and COST Action CA15220. 
All authors gratefully acknowledge support from the COST Action MP1209.


\begin{thebibliography}{99}
\bibitem{remark}
Here, and only within this section, we use the letter $n$ as superscript to label the particle with which the engine interacts, and \emph{not} the excitation number of the harmonic oscillator, as in the rest of the article.
\end{thebibliography}
\end{document}